\documentclass[aps,prl,twocolumn,groupedaddress]{revtex4-1}
\usepackage{graphicx}% Include figure files
\usepackage{dcolumn}% Align table columns on decimal point
\usepackage{bm}% bold math
\begin{document}

%\preprint{}

\title{On the existence of natural self-oscillation of a free electron}


%\thanks{}
\author{Zhixian Yu}
\affiliation{Department of Physics and Astronomy, University of New Mexico, Albuquerque NM 87131 USA}
\affiliation{College of Physics Science, Qingdao University, Qingdao 266071 China}
\author{Liang Yu}
\email{lyuqdcn@gmail.com}
\affiliation{Laboratory of Physics, Jiaming Energy Research, Qingdao 266003 China}

\date{\today}

\begin{abstract}
The possibility of the existence of natural self-oscillation of a
free electron is suggested. This oscillation depends on the
interaction of the electron with its own electromagnetic fields.
Suitable standing wave solutions of the electromagnetic fields are
chosen. A kind of displacement dependent electric potential and
mechanism of energy exchange between velocity and acceleration
dependent electromagnetic fields are analyzed. Conditions for the
existence of natural self-oscillation are given.
\end{abstract}
%\keywords{}

\maketitle
\section{introduction}
The subtle and beautiful experiment of Dehmelt et al showed that a
single isolated electron might be driven by static electric field in
a Penning trap to oscillate as a ``mono-election oscillator"\cite{Wineland}.
We want to show here that a free electron may also possess natural
self-oscillation by itself under some selected conditions.

The property of a free electron has been discussed by many authors
through both classical and quantum theories{\cite{Rohrlich, Coleman}}. There are still
difficulties and paradoxes about the model of electron. Here we take
a simple model of electron used by Konopinsky\cite{Konopinsky} and
Dehmelt{\cite{Dehmelt}}, in which a free electron is a system consisting of a
particle with mass $m$, charge $e$ and distributed electromagnetic
fields. These fields may have their own energies, momentums and
angular momentums. The more detailed internal structures of the
electron are neglected at first, thus many difficulties about the
model of electron are avoided.
\section{Standing wave solutions of electromagnetic fields of an electron in harmonic oscillation}
The electromagnetic fields of an electron in harmonic oscillation
may be derived from the vector potential or Hertz vector of the
oscillating electron{\cite{Marion,Jackson2,Panofsky}}. Following the method of
Bateman\cite{Bateman} or Adler, Chu and Fano{\cite{Adler}}, we may also solve the
wave equation of the electromagnetic fields on the basis of Maxwell
equations in spherical coordinates first, and then match the
solutions with the motion of the electron. From both methods we may
have the dipole term of the electron's electromagnetic fields
as\cite{Marion}
\begin{equation}
{\bm {E_r}}=2eZ_{e0}[{1\over {r^3}}-{ik\over
r^2}]\cos\theta{e^{i\omega{t}}}e^{\mp ikr}\hat{\bm r},
\end{equation}
\begin{equation}
{\bm {E_\theta}}=eZ_{e0}[{1\over {r^3}}-{ik\over r^2}-{k^2\over
r}]\sin\theta{e^{i\omega{t}}}e^{\mp ikr}\hat{\bm \theta},
\end{equation}
\begin{equation}
{\bm {H_\phi}}=eZ_{e0}[-{ik\over r^2}-{k^2\over
r}]{\sin}\theta{e^{i\omega{t}}}e^{\mp ikr}\hat{\bm \phi},
\end{equation}
where ${\bm {E_r}}$ and ${\bm {E_\theta}}$ are electric fields in
$r$ and $\theta$ directions, ${\bm {H_\phi}}$ is magnetic field
in $\phi$ direction, $\hat{\bm r}$, $\hat{\bm \theta}$ and
$\hat{\bm \phi}$ are corresponding unit vectors (Besides Eqs. (1), (2) and (3), there are also a monopole Coulomb field of the electron and a dipole field at the origin{\cite{Jackson2}}, but they are not useful in
following analysis), $r$ is the distance from the electron to a field
point, $e$ is charge of electron, $Z_{e0}$ is amplitude of its
oscillation, terms of $1/r^3$, $1/r^2$ and $1/r$
correspond to oscillating static fields, velocity dependent near
fields and acceleration dependent fields respectively.

The $\mp$ signs in the space phase factor represent that retarded
and advanced time solutions are used respectively. These two are
complex conjugate solutions which represent progressive divergent
wave and regressive convergent wave, and any combination of them is
also an allowed solution. Usually the solution of advanced time is
rejected by the requirement of causality, but here it is used as one of the conjugate waves, there is no problem about causality. According to Adler, Chu and Fano the two kinds of space factors may add together
to form complete standing wave modes\cite{Adler} of $\sin{kr}$ and $\cos{kr}$. We choose the sum of retarded time and advanced time solutions to form standing wave solutions. Then their space and time
phase factors are formed as
\begin{equation}
\small e^{i(\omega{t}-kr)}+e^{i(\omega{t}+kr)}=2(\cos{kr}\cos{\omega{t}}+i\cos{kr}\sin{\omega{t}}),
\end{equation}
thus for the static and acceleration dependent fields, the phase factors are $2\cos{kr}\cos{\omega t}$. For the velocity dependent electromagnetic fields, they are $2\cos{kr}\sin{\omega t}$, since
\begin{eqnarray}
-i2(\cos{kr}\cos{\omega{t}}+i\cos{kr}\sin{\omega{t}})\nonumber\\=2(\cos{kr}\sin{\omega{t}}-i\cos{kr}\cos{\omega{t}}).
\end{eqnarray}

In order to match the motion of the oscillating electron, we have to
choose the static electric fields $E_{sr}$, $E_{s\theta}$ and
acceleration dependent electromagnetic fields $E_{a\theta}$,
$H_{a\phi}$ with their phase factors as
\begin{equation}
E_{sr}={4eZ_{eo}\over r^3}\cos{\theta}\cos{kr}\cos{\omega{t}},
\end{equation}
\begin{equation}
E_{s\theta}={2eZ_{eo}\over r^3}\sin{\theta}\cos{kr}\cos{\omega{t}},
\end{equation}
\begin{equation}
E_{a\theta}=-{2k^2eZ_{eo}\over r}\sin{\theta}\cos{kr}\cos{\omega{t}},
\end{equation}
\begin{equation}
H_{a\phi}=-{2k^2eZ_{eo}\over r}\sin{\theta}\cos{kr}\cos{\omega{t}}.
\end{equation}
The velocity dependent electromagnetic fields $E_{vr}$,
$E_{v\theta}$ and $H_{v\phi}$ with their standing wave phase
factors are chosen as
\begin{equation}
E_{vr}=-{4keZ_{eo}\over r^2}\cos{\theta}\cos{kr}\sin{\omega{t}},
\end{equation}
\begin{equation}
E_{v\theta}=-{2keZ_{eo}\over
r^2}\sin{\theta}\cos{kr}\sin{\omega{t}},
\end{equation}
\begin{equation}
H_{v\phi}=-{2keZ_{eo}\over
r^2}\sin{\theta}\cos{kr}\sin{\omega{t}}.
\end{equation}
\section{Displacement dependent electric potential
and force for self-oscillation of a free electron}
It is known that for any electric field $E$ at a point in free
space, there is a corresponding energy density ${\mathcal E}_d$,
which is
\begin{equation}
{\mathcal E}_d={1\over 8\pi}E^2,
\end{equation}
and the total energy ${\mathcal E}$ of the electric field is the
volume integration of its energy density through whole space, which is
\begin{equation}
{\mathcal E}={1\over 8\pi}\int {E^2}dv.
\end{equation}
Thus for the electrical fields $E_{sr}$ and $E_{s\theta}$ of the static
zone, there is the corresponding energy ${\mathcal E}_s$. As the
time factor is suppressed at first,
\begin{widetext}
\begin{eqnarray}
{\mathcal E}_s={1\over 8\pi}\int({E_{sr}^2}+{E_{s
\theta}^2})dv={1\over 8\pi}\int_{r_{min}}^{r_{max}}[{({4eZ_{eo}\over
r^{3}}\cos \theta\cos kr)^2}+{({2eZ_{eo}\over r^{3}}\sin \theta\cos
kr)^2}]2\pi{r^2}dr\sin{\theta}d\theta,
\end{eqnarray}
\end{widetext}
where $r_{max}$ and $r_{min}$ are upper and lower limits of the integration.
It's reasonable to take the upper limit $r_{max}$ as infinite or the
total length of certain number of standing wave lengths, and the
lower limit $r_{min}$ as $M_sr_0$, where $r_0$ is taken as the
classical radius of electron which is defined as\cite{Feynman}
\begin{equation}
r_0={e^2\over mc^2}=2.82 \times 10^{-13}cm,
\end{equation}
and $M_s$ is an undetermined numerical constant. Taking
$r_{max}=\infty$ and $r_{min}=M_sr_0$ we get the integration of
${\mathcal E}_s$ as

\begin{eqnarray}
{\mathcal E}_s=4{e^2Z_{e0}^{2}}[{1+\cos(2kM_sr_0)\over
6({M_s}{r_0})^3}-{k\sin(2kM_sr_0)\over
6({M_s}{r_0})^2}\nonumber\\-{k^2\cos(2kM_sr_0)\over 3M_sr_0}+{{2\over
3}k^{3}}{\int_{M_sr_0}^{\infty}}{\sin2kr\over r}dr].
\end{eqnarray}
Since $r_0$ and $M_s$$r_0$ are comparatively small, we may use the
approximations
\begin{equation}
\cos (2kM_sr_0){\approx}1,
\end{equation}
\begin{equation}
{\sin(2k{M_s}{r_0}){\approx}2kM_sr_0},
\end{equation}
\begin{equation}
{\int_{M_sr_0}^{\infty}}{\sin2kr\over r}dr{\approx}{\pi\over
2}.
\end{equation}
Then we have
\begin{equation}
{\mathcal E}_s=4{e^2Z_{e0}^2}[{1\over 3(M_sr_0)^3}-{2k^2\over
3M_sr_0}+{{\pi\over 3}k^3}].
\end{equation}
Since $Z_{e0}$ is the amplitude of the harmonic oscillation of the
electric fields, ${\mathcal E}_s$ is a displacement dependent energy
which is proportional to the square of $Z_{e0}$ and may has a kind
of restore force with maximum value $f_{s0}$, which is
\begin{equation}
f_{s0}=-{\partial {\mathcal E}_s\over \partial Z_{e0}}=-8e^2[{1\over
3(M_sr_0)^3}-{2k^2\over 3M_sr_0}+{\pi\over 3}k^3]Z_{e0}.
\end{equation}
This restore force will drive the electron to take harmonic
oscillation and we may take energy ${\mathcal E}_s$ equal the
maximum of the electron's kinetic energy ${\mathcal E}_k$, which is
\begin{equation}
{\mathcal E}_k={1\over 2}mv_0^2={1\over 2}mc^2k^2Z_{e0}^{2},
\end{equation}
since $v_0=ckZ_{e0}$ is the amplitude of the electron's velocity.
Then we have
\begin{equation}
{\mathcal E}_s={\mathcal E}_k,
\end{equation}
\begin{equation}
4e^2Z_{e0}^2[{1\over 3(M_sr_0)^3}-{2k^2\over 3M_sr_0}+{\pi\over
3}k^3]={1\over 2}mc^2k^2 Z_{e0}^2,
\end{equation}
\begin{equation}
{\pi\over 3}k^3-({2\over 3M_sr_0}+{{1\over 8}mc^2\over
{e^2}})k^2+{1\over 3(M_sr_0)^3}=0,
\end{equation}
then by Eq. (16) we get
\begin{equation}
{\pi\over 3}k_s^3-({2\over 3M_sr_0}+{1\over 8r_0})k_s^2+{1\over
3(M_sr_0)^3}=0,
\end{equation}
where $k$ is labeled as $k_{s}$. Eq. (27) gives the numerical
relation of $k_s$ with $M_s$ and $r_0$, and is one of the selection
conditions for the natural self-oscillation of the free electron.
This equation may be modified by some
factors such as the range of the upper limit $r_{max}$, the
relativistic variation of the electron's kinetic energy, but the main feature of Eq. (27) is not affected.
\section{The energy storage and exchange of the
velocity and acceleration dependent electromagnetic fields}
Using the electromagnetic fields $E_{vr}$, $E_{v\theta}$ and
$H_{v\phi}$ of Eqs. (10), (11) and (12), we may calculate the energy of the velocity
dependent field ${\mathcal E}_v$ as
\begin{widetext}
\begin{eqnarray}
{\mathcal E}_v={1\over 8\pi}\int({E_{vr}^2}+{E_{v \theta}^2}+{H_{v
\phi}^2})dv={4{k^2}{e^2}{Z_{e0}^2}\over
8\pi}\int_{r_{min}}^{r_{max}}{\cos^2kr\over r^4}(4\cos^2
\theta+2\sin^2 \theta)2\pi r^2dr\sin{\theta}d\theta.
\end{eqnarray}
\end{widetext}
Taking $r_{max}=\infty$ and $r_{min}=M_vr_0$, where $M_v$ is also an
undetermined constant as the $M_s$ in Eq. (17), we get
\begin{equation}
{\mathcal E}_v={8{k^2}{e^2}Z_{e0}^{2}\over
3}[{1+\cos(2k{M_v}{r_0})\over
{M_v}{r_0}}-2k\int_{M_vr_0}^{\infty}{\sin{2kr}\over r}dr].
\end{equation}
Since $r_0$ and $M_v$$r_0$ are comparatively small, we may use the
approximations
\begin{equation}
\cos (2kM_vr_0){\approx}1,\qquad{\int_{M_vr_0}^{\infty}}{\sin2kr\over r}dr{\approx}{\pi\over
2},
\end{equation}
then we have
\begin{equation}
{\mathcal E}_v={8{k^2}{e^2}Z_{e0}^{2}\over 3}({2\over
{M_v}{r_0}}-k\pi).
\end{equation}

The acceleration dependent energy ${\mathcal E}_a$ may
be calculated from the acceleration dependent electromagnetic fields
${E_{a\theta}}$ and ${H_{a\phi}}$ in $\theta$ and $\phi$
directions. According to Eqs. (8) and (9), ${\mathcal E}_a$ is
\begin{equation}
\begin{array}{lll}
{{\mathcal E}_a}&=&\displaystyle{1\over 8\pi}\int(E_{a\theta}^2+H_{a\phi}^2)dv\\[1.2em]
&=&\displaystyle{4k^4e^2Z_{e0}^2\over 8\pi}\int_{r_{min}}^{r_{max}}{{2\sin^2\theta}\over {r^2}}\cos^2kr2{\pi}{r^2}dr{\sin}{\theta} d {\theta}\\[1.2em]
&=&\displaystyle{4k^4e^2Z_{e0}^2\over {3}}{\left. (r+{\sin2kr\over
2k}) \right|}_{r_{min}}^{r_{max}},
\end{array}
\end{equation}

\noindent where $k=2\pi/\lambda$, $\lambda$ is wave length of the
standing wave. The acceleration dependent energy ${\mathcal E}_a$
contains a number of energy bands which are standing wave spherical shells with equal energy. We take the number of the bands each with width $\lambda/2$
as $N$. The central band is bisected by the electron at its center,
half of its width is ${1\over 2}\cdot{\lambda\over 2}$, thus the
upper limit is $r_{max}=(N+{1\over 2}){\lambda\over 2}$. The lower
limit here may be taken as $r_{min}=0$, since ${M_s}{r_0}$ is small
and here $r_{min}$ is not in the denominator as above. Thus
\begin{equation}
\begin{array}{lll}
{{\mathcal E}_a}&=&\displaystyle{4k^4e^2Z_{e0}^2\over {3}}{\left.
(r+{\sin2kr\over
2k}) \right|}_{r_{min}=0}^{r_{max}=(N+{1\over 2}){\lambda\over 2}}\\[1.2em]
&=&\displaystyle{4k^4e^2Z_{e0}^2\over 3}(N+{1\over 2}){\lambda\over 2}\\[1.2em]
&=&\displaystyle{4k^3e^2Z_{e0}^2\over 3}{(N+{1\over 2}){\pi}}.
\end{array}
\end{equation}

The electric and magnetic fields of ${{\mathcal E}_v}$ or
${{\mathcal E}_a}$ respectively have equal time phase and they can not
exchange energy within ${{\mathcal E}_v}$ or ${{\mathcal E}_a}$ alone, but
the electromagnetic fields of ${{\mathcal E}_v}$ and ${{\mathcal
E}_a}$ are out of phase $\pi/2$ in time, thus ${{\mathcal
E}_v}$ and ${{\mathcal E}_a}$ can exchange their stored energy
between each other. The exchange of energy between velocity and
acceleration dependent fields of an oscillating electric dipole had
been discussed by Mandel\cite{Mandel} and Booker\cite{Booker} and they also
showed that there was energy flow between these two fields. This
kind of energy flow is also discussed in antenna theory. The
condition for the complete energy exchange between ${{\mathcal
E}_v}$ and ${{\mathcal E}_a}$ is
\begin{equation}
{\mathcal E}_v={\mathcal E}_a.
\end{equation}
From Eqs. (31) and (33), we get
\begin{equation}
{8k^2e^2Z_{e0}^2\over 3}[{2\over
{M_vr_0}}-k{\pi}]={4k^3e^2Z_{e0}^2\over 3}(N+{1\over 2}){\pi},
\end{equation}
then we have
\begin{equation}
k_v={4\over{M_vr_0[(N+{1\over 2})\pi+2{\pi}]}},
\end{equation}
here $k$ is labeled as $k_v$. This is a selection condition of $k_v$
with $M_v$ and $r_0$. For the free electron to have any kind of
persisted natural self-oscillation, both conditions of $k_s$ in
Eq. (27) and $k_{v}$ in Eq. (36) should be satisfied at the same time,
that is
\begin{equation}
k_s=k_v,
\end{equation}
which is the combined relation of possible natural self-oscillation.
\section{Short discussion}
Natural self-oscillation is popular in electric circuits, microwave
structure or mechanical system. Standing wave solutions of
electromagnetic waves are also used in antenna theory. Above energy
analyses show that a free electron may also possess natural
self-oscillation through the interaction with its own
electromagnetic fields under certain conditions. These conditions
are determined by the electron's standing wave modes, the number $N$
of its standing wave bands in acceleration dependent fields and two
numerical constants in its energy integrations. For an electron
oscillating in a Penning trap as that in the experiment of Dehmelt
for the ``mono-electron oscillator", there is a strong static
magnetic field along the oscillating direction of the electron to
keep the oscillating electron localized and in stable movement. For
the self-oscillation of a free electron this kind of external
auxiliary is unnecessary, since localization is not a problem for a
free electron. The natural self-oscillation will have some effects
on the interaction of the electron with its environment and thus
could be measured by suitable experiments. We will discuss these in
connection with the stability of natural self-oscillation of a free
electron. Here we only give an energy analysis and suggest that its
existence is possible.
\\

\end{document}